\documentclass[lnicst]{svmultln}
\usepackage{color}
\usepackage{leftidx}
\usepackage{cite}
\usepackage{microtype}
\usepackage{cleveref}
\usepackage{graphicx,amssymb,amstext,amsmath,cases,float}
\usepackage[ruled,vlined,lined,ruled,linesnumbered]{algorithm2e}
\usepackage{makeidx}

\begin{document}

\mainmatter

\title{Reusing Wireless Power Transfer for Backscatter-assisted Cooperation in WPCN}

\titlerunning{ }

\author{Wanran Xu \and Suzhi Bi
 \and Xiaohui Lin \and Juan Wang}
\authorrunning{ }
\tocauthor{Suzhi Bi, Xiaohui Lin, Juan Wang}
\institute{College of Information Engineering, Shenzhen University,\\
Shenzhen,  Guangdong, 518060, China\\
\email{{xuwanran2016,bsz,xhlin,juanwang}@szu.edu.cn}}
\maketitle

\begin{abstract}
This paper studies a novel user cooperation method in a wireless powered communication network (WPCN), where a pair of closely located devices first harvest wireless energy from an energy node (EN) and then use the harvested energy to transmit information to an access point (AP). In particular, we consider the two energy-harvesting users exchanging their messages and then transmitting cooperatively to the AP using space-time block codes. Interestingly, we exploit the short distance between the two users and allow the information exchange to be achieved by energy-conserving backscatter technique. Meanwhile the considered backscatter-assisted method can effectively reuse wireless power transfer for simultaneous information exchange during the energy harvesting phase. Specifically, we maximize the common throughput through optimizing the time allocation on energy and information transmission. Simulation results show that the proposed user cooperation scheme can effectively improve the throughput fairness compared to some representative benchmark methods.
\end{abstract}

\section{Introduction}
Wireless device battery life has always been a key problem in modern wireless communication. Frequent battery replacement/recharging may bring lots of inconvenience and cause high probability of communication interruption. To overcome the above difficulties, RF-enabled wireless energy transfer (WET) technique has recently drawn greater attention \cite{2014:Bi,2016:Bi1,2015:Lu}, which can charge wireless devices with continuous and stable energy through the air.

One useful application of WET is wireless powered communication network (WPCN) \cite{2014:Ju1,2016:Bi2,2016:Bi3,2017:Bi1,2018:Xu,2018:Bi,2015:Bi}, where wireless devices (WDs) transmit information using the energy harvested from energy node. Specifically, \cite{2014:Ju1} proposed a harvest-then-transmit protocol in WPCN where one hybrid access point (HAP) with single-antenna first broadcasts energy to all users, then allows users to take turns to perform wireless information transmission (WIT). \cite{2016:Bi2} studied the placement optimization when each pair of EN and AP is colocated and integrated as a hybrid access point. However, all the above works consider using a HAP for performing both WET and WIT. This WPCN model will inevitable suffered from a so-called ``doubly near-far" problem, such that the far user can receive less wireless power than the near user, but needs to consume more energy to transmit information for achieving the same communication performance.

To enhance user fairness, \cite{2014:Rui} proposed a two-user cooperation scheme where the near user helps relay the far user's information to the HAP. \cite{2017:Yuan} considered a cluster-based user cooperation in a WPCN with a multi-antenna HAP. However, using a single HAP is the essential cause of the user unfairness problem. To further enhance system performance, \cite{2016:Zhong1} considered using separate EN and information AP. Specifically, the WDs first harvest energy from the EN and then use distributed Alamouti code to jointly transmit their information to the AP. Thanks to the achieved cooperative diversity gain, the cooperation scheme can effectively improve the throughput performance. However, the two WDs may need to consume considerable amount of energy and time on information exchange, which may constrain the overall communication performance of the energy-constrained devices.

Besides, a newly emerged low-cost ambient backscatter (AB) communication technique provides an alternative method to reduce such cooperation system's overhead \cite{2015:Backscatter1,2013:Backscatter2,2016:Backscatter3}. Specifically, AB allows WDs to transmit information by passively backscatter environment RF signals, e.g., WiFi and cellular signals, in the neighbouring area. Several recent works have focused on improving the data rates of AB, such as applying new signal detection methods and more advanced backscattering circuit designs \cite{2013:Backscatter2,2016:Backscatter3}. However, due to the dependency on time-varying environment RF signals, backscatter technology suffers many problems, e.g., sensitive to transmission distance and uncontrollable transmissions.

In this paper, we present a novel user cooperation method assisted by backscatter in WPCN. As shown in Fig.~1, we consider a similar setup in \cite{2016:Zhong1}, where two devices first harvest WET from the HAP and then transmit jointly to the AP by forming a virtual antenna array. However, unlike the conventional information exchange in \cite{2016:Zhong1}, we allow the two closely-located WDs to use backscatter communication to exchange their messages in a passive manner during the WET stage. The key contributions of this passage are summarized as follows:

\begin{enumerate}
    \item We propose a novel user cooperation method assisted by backscatter communication during the information exchange stage. By reusing the WET signal, the proposed method can achieve simultaneous energy harvesting and information exchange, and thus potentially improves the throughput performance of the energy-constrained system.
    \item We derive the individual throughput of the two WDs for the proposed backscatter-assisted cooperation method, and formulate an optimization problem that maximizes the minimum data rates (common throughput) between the two users. By optimizing the time allocation on WET and WIT, we can effectively enhance the throughput fairness of the system.
    \item We show that the throughput maximization problem can be cast as a convex optimization, such that its optimum can be efficiently obtained. By comparing with some representative benchmark methods, we show that the proposed backscatter-assisted cooperation can effectively improve the throughput performance under various practical network setups.
\end{enumerate}

\begin{figure}
\vspace{-0.6cm}
\setlength{\abovecaptionskip}{0pt}
\setlength{\belowcaptionskip}{10pt}
  \centering
   \begin{center}
      \includegraphics[width=0.6\textwidth]{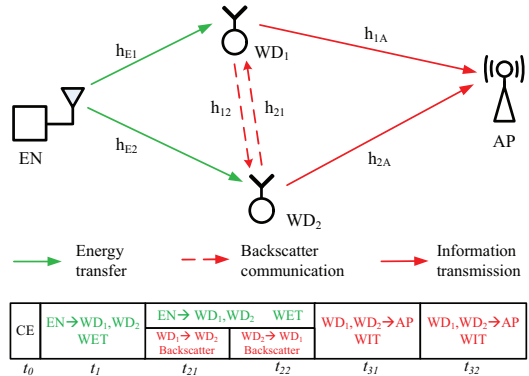}
   \end{center}
  \caption{The proposed user cooperation method and operating protocol.}
  \label{Fig.4}
\end{figure}

\section{System Model}

\subsection{Channel Model}
As shown in Fig.~1, we consider a WPCN consisting of an EN, two WDs, and an information AP, where all the devices have single antenna each. The EN is assumed to have stable energy supply and able to broadcast RF energy at constant power $P_0$. Besides, it has a time-division-duplexing (TDD) circuit structure to switch between energy transfer and communication, e.g, for performing channel estimation. The two WDs have no other embedded energy source thus need to harvest RF energy for performing information transmission to the AP.

The circuit block diagram of a WD is shown in Fig.~2. With the two switches $S_1$ and $S_2$, a WD can switch flexibly among three operating modes as follows.

\begin{figure}
\vspace{-0.4cm}
\setlength{\abovecaptionskip}{0pt}
\setlength{\belowcaptionskip}{10pt}
\centering
  \begin{center}
    \includegraphics[width=0.6\textwidth]{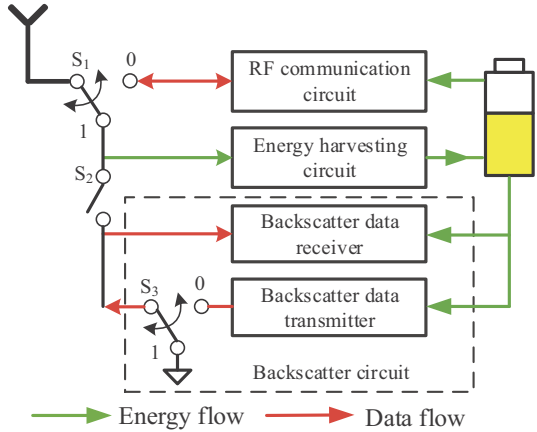}
  \end{center}
  \caption{Circuit block diagram of the RF-powered backscatter wireless device.}
  \label{102}
\end{figure}

\begin{enumerate}
    \item \emph {RF (active) communication mode} ($S_1 = 0$): the antenna is connected to the RF communication circuit and WD is able to transmit or receive information using conventional RF wireless communication techniques, e.g., QAM modulation and coherent detection.
    \item \emph {energy harvesting mode} (S$_1=1$ and S$_2$ is open): the antenna is connected to the energy harvesting circuit, which can convert the received RF signal to DC energy and store in a rechargeable battery. The energy is used to power the operations of all the other circuits.
    \item \emph{backscatter (passive) communication mode} (S$_1$=1 and S$_2$ is closed): the antenna is connected to backscatter communication and energy harvesting circuits. In this case, the WD transmits information passively by backscattering the received signal. Specifically, by setting the switch S$_3=0$, the impedance-matching circuit absorbs most of the received signal such that a ``0" is transmitted; otherwise when S$_3=1$, due to the imbalance of transmission line impedance, the received signal is reflected and broadcasted by the antenna such that a ``1" is transmitted. Meanwhile, non-coherent detection techniques, e.g., energy detector \cite{2012:Dector}, can be used to decode backscatter transmissions from other devices.
\end{enumerate}

Notice that a WD can harvest RF energy simultaneously when the backscatter circuit transmits or receives information. Specifically, as shown in Fig.~3, a power splitter is used to split the received RF signal into two parts. We denote the portion of signal power for backscatter communication by $(1-\beta)$, where $\beta\in[0,1]$, and the rest $\beta$ for energy harvesting (EH). The received signal is corrupted by an additive noise $N_0 \sim \mathcal{CN}(0,\sigma_0^{2})$ at the receiver antenna. Besides, the power splitting circuit and the information decoding circuit are also introduced by an additional noise $N_s \sim \mathcal{CN}(0,\sigma_s^{2})$, which is assumed independent of the antenna noise $N_0$. As a result, the equivalent noise power for information decoding is $(1- \beta) \sigma_0^2 + \sigma_s^2$. The value of $\beta$ can be adjusted according to different receive signal power and is assumed constant for the time being.

\begin{figure}
\vspace{-0.6cm}
\setlength{\abovecaptionskip}{0pt}
\setlength{\belowcaptionskip}{10pt}
\centering
  \begin{center}
    \includegraphics[width=0.7\textwidth]{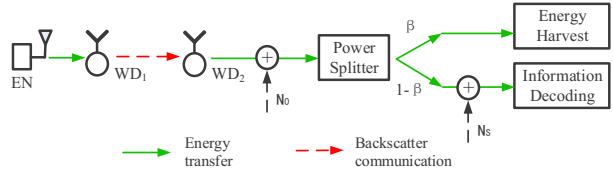}
  \end{center}
  \caption{The power splitter structure used during backscatter communication}
  \label{103}
\end{figure}

Evidently, backscatter communication does not need to generate RF carrier signals locally and insteads using simple energy encoding/decoding circuits, thus is more energy-efficient than conventional active wireless communication. However, its application is often limited by the strength of the ambient RF signal and its short communication range (within a couple of meters) due to the weak signal strength after reflection. In the following, we propose a method to reuse WET to achieve controllable backscatter communication in WPCN.
\subsection{Protocol Description}
We consider a block fading channel model where all the channels are reciprocal and the channel gains remain constant during each transmission block of duration $T$. At the beginning of a transmission block channel estimation (CE) is performed within a fixed duration $t_0$. Then, a three-stage operating protocol is used in the remainder of a tagged transmission block, as shown in Fig.~1. Specifically, in the second stage, the EN continuous to broadcast energy for $t_2$ amount of time, during which the WD$_1$ and WD$_2$ take turns to backscatter their local information for $t_{21}$ and $t_{22}$ amount of time, respectively, where $t_2 = t_{21} + t_{22}$. With the power splitter structure in Fig.~3, each of the two WDs can decode information from the other's transmission and harvest RF energy simultaneously. Notice that we neglect the backscatter signal received by the AP due to the much larger distance separation between a WD and the AP in practice. In the third stage of length $t_3$, the two users transmit jointly their information to the AP. Specially, $t_{31}$ amount of time is allocated to transmit user WD$_1$'s information, and the rest of $t_{32}$ is for transmitting WD$_2$'s information. Accordingly, we have a total time constraint
\begin{equation}
\label{1}
t_0 + t_1 + t_{21} + t_{22} +t_{31} +t_{32} = T.
\end{equation}
For convenience, we normalize $T$ = 1 in the sequel without loss of generality.

We denote the complex channel coefficient between WD$_1$ and WD$_2$ as $\alpha_{12}$. Similarly, the other channel coefficients are denoted as $\alpha_{E1}$, $\alpha_{E2}$, $\alpha_{1A}$, $\alpha_{2A}$, $\alpha_{21}$. In the CE stage, user WD$_1$ and WD$_2$ broadcast their pilot signals, so that EN has the knowledge of $\alpha_{E1}$ and $\alpha_{E2}$, the AP knows $\alpha_{1A}$ and $\alpha_{2A}$, and user WD$_1$ (WD$_2$) knows $\alpha_{12}$ ($\alpha_{21}$) respectively. Then, each node feeds back their known CSI to a control point, which calculates and broadcasts the optimal time allocation ($t_1^*$,$t_{21}^*$,$t_{22}^*$,$t_{31}^*$,$t_{32}^*$) to all the nodes in the network.

\section{System Performance Analysis}
\subsection{Derivation of Individual Data Rate}
During the WET phase, we denote the baseband equivalent pseudo-random energy signal transmitted by the EN as $x(t)$ with $E[|x(t)|^2] = 1$. The received signal at WD$_i$, $i = 1, 2$, is then expressed as
\begin{equation}
\label{2}
y_i^{(1)} (t) =\sqrt{P_0} \alpha_{Ei} x(t) + n_i(t),
\end{equation}
where $n_i(t)$ denotes the receiver noise at WD$_i$ with $n_i(t) \sim \mathcal{CN}(0,N_0)$. It is assumed that $P_0$ is sufficiently large such that the energy harvested due to the receiver noise is negligible. Hence, the amount of energy harvested by WD$_1$ and WD$_2$ can be expressed as
\begin{equation}
\label{3}
E_1^{(1)} = P_0 \eta h_{E1} t_1, ~~~~ E_2^{(1)} = P_0 \eta h_{E2} t_1,
\end{equation}
where $0 < \eta < 1$ denotes the energy harvesting efficiency coefficient.

In the backscattering stage, WD$_1$ first transmits its information to WD$_2$ for $t_{21}$ amount of time. We assume a fixed data transmission rate $R_b$ bit/s, thus the duration of transmitting a bit is $1/R_b$ second. In particular, when WD$_1$ transmits a bit $0$, the switch $3$ is open, and WD$_2$ receives only the energy signal from the HAP and WD$_1$. Otherwise, when WD$_1$ transmits a bit $1$, the received signal at WD$_2$ is a combination of both the HAP's energy and the reflected signal from WD$_1$. Those signals can be jointly expressed as
\begin{equation}
y_{2}^{(2)}(t) = \alpha_{E2} \sqrt{P_0} x(t) + B \mu_1 \alpha_{E1} \alpha_{12} \sqrt{P_0} x(t) + n_2^{(2)}(t),
\end{equation}
where $\mu_1$ denotes the backscatter reflection coefficient of WD$_1$, and $B$ denotes the information bit transmitted by WD$_1$ through backscattering. Due to the use of the power splitter at each user, the energy and information signals received by WD$_2$ can be respectively expressed as
\begin{equation}
y_{2,E}^{(2)}(t) = \sqrt{\beta} y_2^{(2)}(t), ~~ y_{2,I}^{(2)}(t) = \sqrt{(1-\beta)} y_2^{(2)}(t).
\end{equation}
It is assumed that the probabilities of transmitting 0 and 1 are equal. Therefore the harvested energy by WD$_2$ can be expressed as
\begin{equation}
\begin{split}
E_2^{(2)} &= \frac{1}{2} \eta \beta t_{21} (E[|y_{2,0}^{(2)}(t)|^2] + E[|y_{2,1}^{(2)}(t)|^2]) \\
          &= \eta \beta t_{21} P_0 [h_{E2} + \mu_1 \alpha_{E1} \alpha_{E2} \alpha_{12} + \frac{1}{2} \mu_1^2 h_{E1} h_{12}].
\end{split}
\end{equation}

Notice in (4), we assume that the signals received directly from the HAP and that reflected from WD$_1$ are uncorrelated due to the random phase change during backscatter. We denote the sampling rate of WD$_2$'s backscatter receiver as $S$, such that it sampled $N = \frac{S}{R_b}$ samples during the transmission of a bit information, where $R_b$ denotes the fixed backscatter rate in bits per second. In the following lemma, we derive the bit error rate (BER) of a backscatter receiver using an optimal energy detector.

\underline{\emph{Lemma}} \emph{3.1}:
The BER of WD$_2$ for the considered backscatter communication with an optimal energy detector is
\begin{equation}
\begin{split}
P_{e2} = \frac{1}{2} erfc \left[\frac{(1-\beta) P_0 \sqrt{N}}{4(1-\beta) \sigma_0^2 + 4 \sigma_s^2} (\mu_1^2 h_{E1} h_{12})\right].
\end{split}
\end{equation}

\emph{Proof}: The proof is omitted here due to the space limitation.

As the backscatter communication can be modeled as a binary symmetric channel, the channel capacity (in bit per channel use) of the transmission from WD$_1$ to WD$_2$ can be expressed as
\begin{equation}
C_2 = 1 + (1 - P_{e2}) \log_2{(1 - P_{e2})} + P_{e2} \log_2{(P_{e2})}.
\end{equation}
By symmetry, we can get the BER and channel capacity from WD$_2$ to WD$_1$ as $P_{e1}$ and $C_1$, where
\begin{equation}
\begin{split}
P_{e1} = \frac{1}{2} erfc \left[\frac{(1-\beta) P_0 \sqrt{N}}{4(1-\beta) \sigma_0^2 + 4 \sigma_s^2} (\mu_2^2 h_{E2} h_{21})\right],
\end{split}
\end{equation}
\begin{equation}
C_1 = 1 + (1 - P_{e1}) \log_2{(1 - P_{e1})} + P_{e1} \log_2{(P_{e1})}.
\end{equation}
As a result, the communication rates of WD$_1$ and WD$_2$ in this stage can be expressed as function of time allocation $\mathbf{t}=[t_0,t_1,t_{21},t_{22},t_{31},t_{32}]$
\begin{equation}
R_1^{(2)}(\mathbf{t}) = R_b t_{21} C_2, ~~ R_2^{(2)}(\mathbf{t}) = R_b t_{22} C_1.
\end{equation}

In the last WIT stage, we assume that both user WD$_1$ and WD$_2$ exhaust the harvested energy, and each transmits with a constant power. Then, the transmit powers of WD$_1$ and WD$_2$ is
\begin{equation}
P_1 = \frac{E_1^{(1)} + E_1^{(2)}}{t_3}, ~~ P_2 = \frac{E_2^{(1)} + E_2^{(2)}}{t_3},
\end{equation}
where $t_3 = t_{31} + t_{32}$. In this stage, the two users use Alamouti STBC transmit diversity scheme \cite{2016:Zhong1} for joint information transmission with $t_{31}$ = $t_{32}$, where the achievable data rates from user WD$_1$ to AP is
\begin{equation}
R_1^{(3)}(\mathbf{t}) = \frac{t_3}{2} \log_2(1 + \frac{P_1 h_{1A}}{\sigma_0^2} + \frac{P_2 h_{2A}}{\sigma_0^2}).
\end{equation}
Likewise, we have $R_1^{(3)}(\mathbf{t}) = R_2^{(3)}(\mathbf{t})$ for user WD$_2$.

\subsection{Common Throughput Maximization}
With the considered cooperation scheme, the overall achievable date rates of user WD$_1$ and WD$_2$ are
\begin{equation}
R_1(\mathbf{t}) = \min\left\{R_1^{(2)}(\mathbf{t}) ,R_1^{(3)}(\mathbf{t})\right\}, ~~
R_2(\mathbf{t}) = \min\left\{R_2^{(2)}(\mathbf{t}) ,R_2^{(3)}(\mathbf{t})\right\}.
\end{equation}

In this paper, we focus on maximizing the common throughput (max-min throughput) of two users by jointly optimizing the time allocated to the HAP, WD$_1$ and WD$_2$.
\begin{equation}
   \begin{aligned}
(P1):\quad & \max_{\mathbf{t}} & & \min\left(R_1(\mathbf{t}),R_2(\mathbf{t})\right) \\
     &\text{s. t.}
     & &   t_0 + t_1 + t_{21} + t_{22} +t_{31} +t_{32} = 1, \\
     & & & t_1,t_{21},t_{22},t_{31},t_{32} \geq 0.
    \end{aligned}
\end{equation}
By introducing an auxiliary variable $Z$, (P1) can be equivalently written as
\begin{equation}
   \begin{aligned}
(P2):\quad & \max_{\mathbf{t},Z} & & Z \\
     &\text{s. t.}
     & &   t_0 + t_1 + t_{21} + t_{22} +t_{31} +t_{32} = 1, \\
     & & & t_1,t_{21},t_{22},t_{31},t_{32} \geq 0, \\
     & & & Z \leq R_1^{(2)}(\mathbf{t}), Z \leq R_2^{(2)}(\mathbf{t}),  \\
     & & & Z \leq {R_1^{(3)}(\mathbf{t}), Z \leq R_2^{(3)}(\mathbf{t}).}
    \end{aligned}
\end{equation}
Notice that $R_1^{(3)}(\mathbf{t}), R_2^{(3)}(\mathbf{t})$ are both concave functions, therefore (P2) is a convex problem whose optimum can be efficiently solved using off-the-shelf algorithms, e.g., interior point method.

\section{Simulation Results}
In this section, we evaluate the performance of the proposed backscatter-assisted cooperation with that without backscatter in \cite{2014:Rui} (the No B.S. scheme). Unless otherwise stated, it is assumed that users are separated by 4 meters. The noise power $N_0$ is set $10^{-10}$W for all receivers, and the additional noise power for ID circuit is $N_s = 10^{-10}$W. The transmit power of EN is $P_0$ = 1W, and the wireless channel gain $h_{ij} = G_A (\frac{3*10^8}{4 \pi d f_d})^{\lambda}$, where $ij \subset\{E1; E2; 1A; 2A; 12; 21\}$, $f_d$ denotes 915 MHz carrier frequency, $\lambda = 2.5$ denotes the path loss exponent, and we fix the antenna power gain $GA = 2$, the signal bandwidth is $10^5$ Hz, and the sampling rate $S=6 \times 10^5$. Without loss of generality, we assume the power splitting factor $\beta = 0.7$, energy harvesting efficiency $\eta = 0.8$, and backscatter reflection coefficient $\mu_1 = \mu_2 = 0.8$.

\begin{figure}
\vspace{-0.6cm}
\setlength{\abovecaptionskip}{0pt}
\setlength{\belowcaptionskip}{10pt}
\centering
  \begin{center}
    \includegraphics[width=0.7\textwidth]{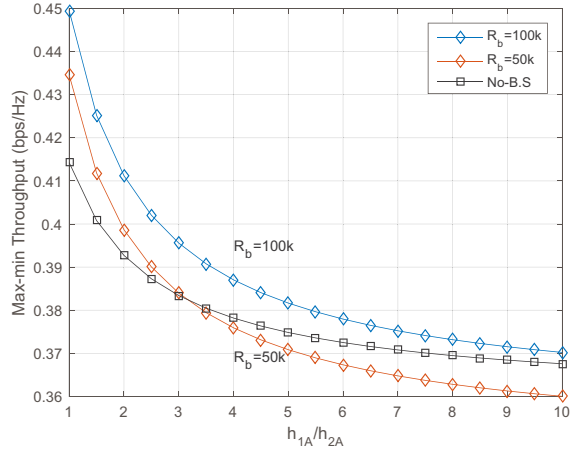}
  \end{center}
  \caption{The impact of user-to-AP channel disparity to the common throughput performance}
  \label{105}
\end{figure}

Fig.~4 shows the impact of user-to-AP channel disparity to the optimal common throughput performance. Here we set $h_{E1} = h_{E2} = 8.5 \times 10^{-5}$, fix $h_{1A} = 8.5 \times 10^{-6}$ as a constant and show the performance when $h_{2A}$ becomes smaller. Notice that when $h_{1A}/h_{2A}$ changes from 1 to 10, all the schemes show a decreasing trend in system performance, which is due to the weaker user-to-AP channel. In particular, the backscatter communication rate $R_b$ has a significantly effect on system performance. As we can see in Fig.~4, when $R_b=50$ kbps the performance of backscatter system is similar to the case without backscattering. However, the former decrease faster than the latter, which is because the worse channel $h_{2A}$ will affect both two user's communication rate. When $R_b$ increases to $100$ kbps, the system performance increases and outperforms the one without backscattering in all cases. This is because the higher backscatter communication rate can effectively reduce the time spent on cooperation, thus leaving more time on energy harvesting and information transmission to the AP.

\begin{figure}
\vspace{-0.6cm}
\setlength{\abovecaptionskip}{0pt}
\setlength{\belowcaptionskip}{10pt}
\centering
  \begin{center}
    \includegraphics[width=0.7\textwidth]{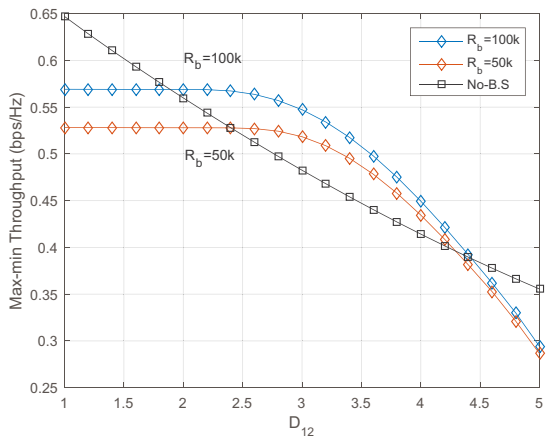}
  \end{center}
  \caption{The impact of inter-user channel to the common throughput performance}
  \label{105}
\end{figure}

Fig.~5 further studies the impact of inter-user channel strength to the throughput performance. Here we set $h_{E1} = h_{E2} = 8.5 \times 10^{-5}$, and $h_{1A} = h_{2A} = 8.5 \times 10^{-6}$. We consider the distance between WD$_1$ and WD$_2$ varies form 1 to 5m. It is observed that the max-min throughput of all schemes decreases with $D_{12}$, due to the worse inter-user channel $h_{12}$. We can see that cooperation without backscatter performs relatively well when $d_{12}$ is small. However, as the distance between users increases, its performance quickly degrades due to the larger time and energy consumed on information exchange, and in general is worse than the proposed backscatter-assisted method, e.g., when $2<d<4.4$ for $R_b=100$ kbps. When the inter-user distance becomes very large, e.g., larger than $4.4$ meters, the common throughputs of the backscatter-assisted cases decrease faster than the case without backscatter because of the extremely sensitivity of backscatter technique to distances. We can therefore conclude that proposed backscatter-assisted cooperation has advantage over that without backscattering when the inter-user channel is relatively weak.

\section{Conclusion}
This paper studied a novel user cooperation method in a two-user WPCN assisted by backscatter communication. In particular, the considered backscatter-assisted method reuses wireless power transfer for simultaneous information exchange during the energy harvesting phase, which can effectively save the energy and time consumed by conventional active transmission schemes. We derived the maximum common throughput of the proposed method through optimizing the time allocation on WET and WIT. By comparing with existing benchmark method, we showed that the proposed method can effectively improve the throughput fairness performance under various practical network setups.

\section*{Acknowledgement}
The work of S. Bi was supported in part by the National Natural Science Foundation of China under Project 61501303, the Foundation of Shenzhen City under Project JCYJ20160307153818306 and JCYJ20170818101824392, the Science and Technology Innovation Commission of Shenzhen under Project 827/000212, and the Department of Education of Guangdong Province under Project 2017KTSCX163. The work of X. H. Lin was supported by research Grant from Guangdong Natural Science Foundation under the Project number 2015A030313552. X. H. Lin is the corresponding author of this paper.

\end{document}